\documentstyle[eqsecnum,epsf,aps, multicol]{revtex}
\begin{document}
\bibliographystyle{prsty}
\draft

\title{Correlation Effects of Single-Wall Carbon Nanotubes in Weak 
Coupling}
\author{Hsiu-Hau Lin}
\address{Department of Physics, University of 
              California, Santa Barbara, CA 93106-9530}
\date{\today}
\maketitle

\begin{abstract}
Single-wall carbon nanotubes with on-site interaction $U \ll t$ are 
studied by a controlled renormalization group method.  When 
formulated as a system of $N_{f}$ flavors of interacting Dirac 
fermions, the effective model at generic filling resembles a set of 
quasi-1D Hubbard chains. For the undoped $(N_{y}, 
N_{y})$-armchair and the $(N_{x}, -N_{x})$-zigzag with $N_x=3n$, 
specific mappings to the 2-chain Hubbard model {\em at half filling} 
with effective interaction $u_{{\rm eff}}=U/N_{y}$ and $2U/N_{x}$ are 
found from this more general equivalence.  The phase diagrams of the 
(3, 3)-armchair and the (6, -6)-zigzag at generic fillings are studied 
and compared as an example.
\end{abstract}

\begin{multicols}{2}

\section{Introduction}
Carbon nanotubes consist of single or mutiple seamless shells of 
graphite sheets with a diameter of about several nanometers and length 
up to micron scale\cite{Ebbesen96a}. The nanotubes reveal rich 
varieties of transport properties as metals, semi-metals and 
semiconductors depending on their radius and 
helicity\cite{Hamada92,Mintmire92,Saito92}. Their marvelous mechanical 
and electrical properties have attracted lots of attention in both 
theory and experiment since their discovery in 1991\cite{Iijima91}. 
Recent measurements show that crystalline ropes of single-wall 
nanotubes (SWNT) exhibit metallic transport 
properties\cite{Thess96,Fischer97,Bockrath97}. An individual SWNT has 
been sucessfully attached to leads in a two-terminal measurement and 
has indicated clear signatures of a 1D quantum wire\cite{Tans97}. The 
quasi-1D nature of these nanotubes necessitates the study of 
interaction effect, which has pronounced influences in low dimensions.  
A lot of theoretical work focuses on the band structure 
calculations\cite{Hamada92,Mintmire92,Saito92,Blase94,Kane97}. The 
$(N, N)$-armchair tube with an on-site interaction $U$ has been mapped 
to the 2-chain Hubbard model with a reduced interaction strength 
$u_{{\rm eff}}=U/N$ {\em at or near} half filling by L.  Balents and 
M.  P.  A.  Fisher\cite{Balents97}. Similar approaches with more 
general interactions are studied independently by Y.  Krotov, D.  H.  
Lee, and S.  Louie\cite{Krotov96u}.  The correlation effect at generic 
filling has been addressed so far.  In this paper, we study a SWNT at 
generic filling with an on-site interaction, which turns out similar 
to the Hubbard chains.

A SWNT is made by rolling up a 2D graphite sheet along the direction 
of a superlattice vector $(N,M)$ and then identifying all sites 
related by this vector (for example, see Fig.  1).  In this paper, we 
focus on two specific types of SWNT's, the $(N,N)$-armchair and the 
$(N,-N)$-zigzag.  There are equivalent wrappings due to the symmetry 
of a honeycomb lattice, for example, $(N,0)$ and $(0,N)$ superlattice 
vectors also produce zigzag tubes.  In the weak coupling limit $U \ll 
t/N, t_{\perp}/N$, it is natural to diagonalize the hoppings and focus 
on the low-energy parts of the whole band structure first. 
The chemical potential cuts through the 
bands at $N_{f}$ pairs of Fermi points, which enable us to formulate 
the low-energy theory as $N_{f}$ flavors of interacting Dirac 
fermions.  The Fermi velocities of these fermions can be computed 
directly from the band structure.  Finding all four-fermion vertices 
generated by the on-site interaction proves to be trickier.  Because 
of the on-site character, the resulting vertices are independent of the 
underlying lattice structure and are shown similar to the Hubbard 
chains.  To be precise, interactions of a $(N_{y},N_{y})$-armchair tube 
at the filling, where $N_{f}$ pairs of Fermi points are present, is 
shown equivalent to the $2N_{y}$-chain Hubbard model (with periodic 
boundary condition) in the $N_{f}$ partially filled band region.  
Similarly, interactions of a $(N_{x},-N_{x})$-zigzag tube are the same 
as in the $N_{x}$-chain Hubbard model except for the absence of the 
umklapp interactions in the $k_{y}$ direction (see below).  Knowing 
the Fermi velocities and initial couplings of interactions, the phase 
diagram is determined through the renormalization group approach 
developed previously in Ref.\onlinecite{Balents96,Lin97u}.

Since the most interesting (feasible) filling is at or near half 
filling, specific mappings for nanotubes to two-chain systems based on 
this more general equivalence are studied.  We found that the {\em 
undoped} $(N_{y},N_{y})$-armchair and the $(N_{x},-N_{x})$-zigzag with 
$N_x=3n$ are equivalent to the two-chain Hubbard model {\em at half 
filling} with effective interaction $u_{{\rm eff}}=U/N_{y}$ and 
$2U/N_{x}$ respectively.  Unlike the previously mentioned $2N_{y}$- 
and $N_{x}$-chain Hubbard model mappings, which only consider 
interactions, the mappings to the two-chain models takes into account 
interactions as well as Fermi velocities (band structure).  In fact, 
detailed calculations show that couplings of the $N$-chain Hubbard 
with an on-site interaction $U$ in the $N_{f}=2$ region are {\em 
identically} the same as the 2-chain model with reduced coupling 
strength $2U/N$ (see below).  The possibility of finding the mappings 
to the 2-chain model relies crucially on the fact that the armchair 
and the zigzag ($N_{x}=3n$) have the same Fermi velocities (band 
structure) as the real 2-chain model at half filling.

The phase diagrams of the (3,3)-armchair and the (6,-6)-zigzag 
both of which resemble a 6-chain system, are analyzed by the RG 
approach. So far, we haven't addressed the point that these tubes in 
real life are made up of ``curved'' graphite sheets.
While this curvature effect does 
not change the low-energy excitations of armchair tubes, it 
creates a minor gap $\Delta_{cur} \sim t/N^{2}$ in zigzag tubes even 
without interactions (see below). If the interaction $U$ is much 
smaller than this minor gap, the (6,-6)-zigzag tube is simply an 
insulator. In the following analysis, we focus on the more interesting 
and non-trivial limit where the interaction $U$
is larger than the minor gap caused by the curvature effect. 
Under this assumption, the phase diagrams of the armchairs and the 
zigzags can be compared at different fillings.
At half filling, both of them acquire charge and spin gaps 
caused by interactions and thus are expected to be 
Mott's insulators at zero temperature instead of metals predicted by 
simple band calculation. Upon slight doping into the $N_{f}=2$ region, 
both develop a d-wave quasi-long ranged correlation
of pair fields. It is a one-dimensional analogy of d-wave 
superconductors (SC's). 
The numerical results of the RG analysis indicate that the armchair has 
a higher critical temperature than the zigzag. 
Larger doping makes the armchair (zigzag) 
enter $N_f$=6 (4) regime where superconductivity correlations are destroyed.

The paper is organized as following.  In Sec II, nanotubes at or near 
half filling, i.e.  $N_{f}=2$, are studied and mapped to equivalent 
2-chain models.  In Sec III, nanotubes at generic fillings are 
considered.  By formulating into $N_{f}$ flavors of interacting Dirac 
fermions, the equivalence of couplings for nanotubes and the Hubbard 
ladders is established.  In the last section, implications of RG 
analysis and other theoretical approaches are discussed.  Long-range 
interaction effects and open questions are also discussed, followed by 
a brief summary of the main conclusions.

\section{Nanotubes at or near half filling}
Following many authors\cite{Hamada92,Mintmire92,Saito92},
we consider a 2D graphite sheet first and then 
take account of finite radius and curvature effects of real tubes.  A 
single sheet of graphite is composed of carbon atoms on the sites of a 
honeycomb lattice which is a triangle Bravais lattice with two carbon 
atoms separated by distance $d$ in each unit cell as shown in Fig.  1.  
The $sp_2$ electrons of each atom form three bonds with their neighbors 
and the remaining single $p_z$ electron can tunnel around.  Since the 
honeycomb lattice is bipartite, the nearest neighbor hoppings can be 
viewed as tunneling alternatively between two triangle sublattices.  
A tight-binding model at half filling (one electron per site) is 
introduced to describe electron hoppings in a graphite sheet
\begin{eqnarray}
H_{0}=\!\!\sum_{{\bf r}\in {\bf R},\alpha}\!\! \bigg\{
&-&t c^{\dag}_{1\alpha}({\bf r})c^{}_{2\alpha}
({\bf r}+{\bf a}_{+}+{\bf d})
\nonumber\\
&-&t c^{\dag}_{1\alpha}({\bf r})c^{}_{2\alpha}
({\bf r}+{\bf a}_{-}+{\bf d})
\nonumber\\
&-&t_{\perp}c^{\dag}_{1\alpha}({\bf r})c^{}_{2\alpha}({\bf r}+{\bf d})
+ {\rm h.c.} \bigg\},
\label{kinetic}
\end{eqnarray}
where ${\bf R}=n_{+}{\bf a}_{+}+n_{-}{\bf a}_{-}$ is the lattice vector of a 
triangle lattice with bases ${\bf a}_{\pm}=a(\pm1/2, \sqrt{3}/2)$.  
The displacement vector between these two 
triangle sublattices is ${\bf d}=a(0,-1/\sqrt{3})$ and $c_{i} 
(c^{\dag}_{i})$ is the fermion annihilation (creation) operator on the 
corresponding sublattices.  The interaction effect is included by 
considering on-site repulsion between electrons
\begin{equation}
H_{U}=U \sum_{{\bf r}\in {\bf R},i}
:n_{i\uparrow}({\bf r}) n_{i\downarrow}({\bf r}):,
\label{on_site}
\end{equation} 
where $n_{i\alpha} \equiv c^{\dag}_{i\alpha}c^{}_{i\alpha}$ denotes 
electron density on sublattice $i$ with spin $\alpha$.  Throughout 
this paper, only repulsive interaction ($U>0$) is considered.

\begin{figure}[hbt]
\epsfxsize=3.5in\epsfbox{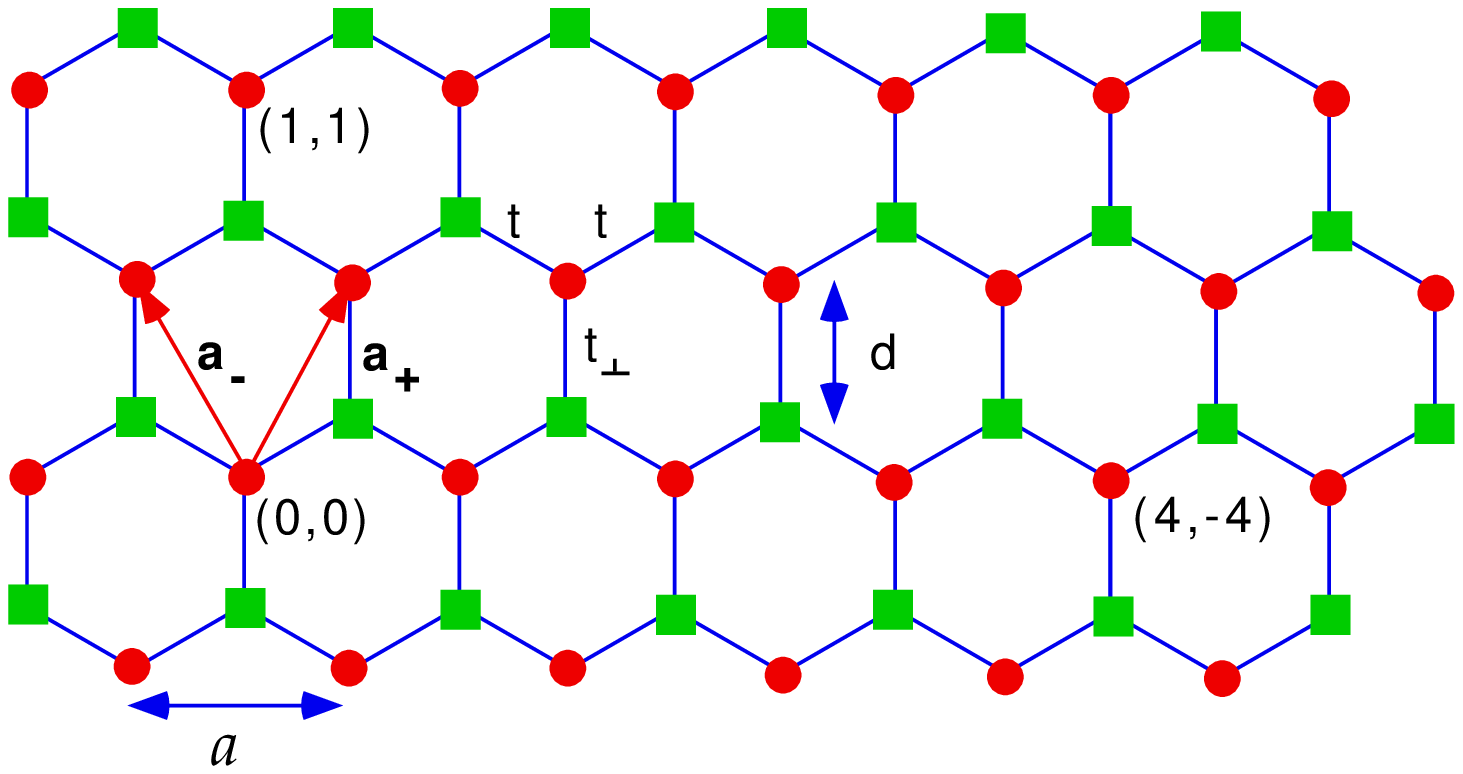}
{\noindent FIG. 1: two dimensional Honeycomb lattice with bases 
${\bf a}_{\pm}$. Circles and squares represent fermion 
sites on two sublattices separated by a distance $d=a/\sqrt{3}$. A 
(1,1)-armchair is obtained by rolling up the sheet from (0,0) to (1,1) 
and end points. Similarly, a 
(4,-4)-zigzag is obtained by rolling from (0,0) to (4,-4).}
\end{figure}

In the weak coupling regime $U \ll t, t_{\perp}$, it is natural to 
diagonalize the hopping Hamiltonian $H_0$ first and study the band 
structure.  By going into momentum space, the energy spectrum is 
$E({\bf k})= \mp |h({\bf k})|$
\begin{equation}
h({\bf k})= 2t \cos(k_x a/2) e^{i k_y a /2\!\sqrt{3}}
+ t_{\perp} e^{-ik_y a/ \sqrt{3}},
\label{band}
\end{equation}
where ${\bf k}$ is the crystal momentum.  The $\mp$ signs in the 
energy spectrum denote bonding and antibonding bands.  Since the 
undoped graphite sheet is half-filled, the bonding band is completely 
filled while the antibonding band is empty. One prominent feature of 
the band structure is that the bonding and antibonding bands are 
separated by a finite gap everywhere in the the Brillouin zone (BZ) 
except at two peculiar points ${\bf K}_{\pm}$.  They are 
conventionally called Dirac points because the (gapless) dispersion 
looks relativistic near them, i.e.
\begin{equation}
E({\bf k}) = v | {\bf k}-{\bf K}_{\pm}|.
\end{equation}
For $t_{\perp}=t$, the Dirac points are located at ${\bf K}_{\pm} 
=(\pm 4\pi/3a)$ and the Fermi velocity $v=(\sqrt{3}/2)ta$ as shown in 
Fig.  2.  If $t_{\perp}$ is slightly different from $t$, these points 
get shifted a bit along the $k_x$ direction.  Equivalent points in the 
BZ are related by reciprocal lattice vectors.  At half-filling (or 
under slight doping), gapless excitations only exist at (or near) the 
Dirac points and the low-energy physics is captured by two flavors of 
Dirac fermions with the on-site interaction in Eq.~\ref{on_site}.  It 
is strikingly similar to the 2-chain Hubbard model which describes two 
interacting Dirac fermions with on-site repulsions.  This
motivates us to search for equivalent 2-chain models for nanotubes near 
half filling.  In the following subsections, we manipulate the $(N_y, 
N_y)$-armchair and the $(N_x,-N_{x})$-zigzag with $N_x=3n$ to show 
that it is indeed possible to map them into equivalent 2-chain models.

\subsection{Armchair tubes}
The $(N_{y}, N_{y})$-armchair has a finite circumference $C=\sqrt{3}a 
N_{y}$ which leads to the quantization of transverse momentum $k_{y}$ 
around the tube.  In addition to the quantized momenta, the curvature 
effect causes the slight difference between $t, t_{\perp}$ which 
shifts the locations of the Dirac points as shown in Fig.  2.  Since 
the quantized momentum ($k_{y}=0$) cuts through the Dirac points, the 
armchair without interactions is expected to be a metal.  But once the 
interaction effect is taken into consideration, the low-energy 
excitations change dramatically.  An armchair carbon nanotube with 
on-site interaction has been shown equivalent to the 2-chain Hubbard 
model at or near half filling where only $k_{y}=0$ band is 
gapless\cite{Balents97}. Here we rederive the same mapping by a more 
systematic method which can also be applied to zigzag tubes later.

\begin{figure}[hbt]
\epsfxsize=3.5in\epsfbox{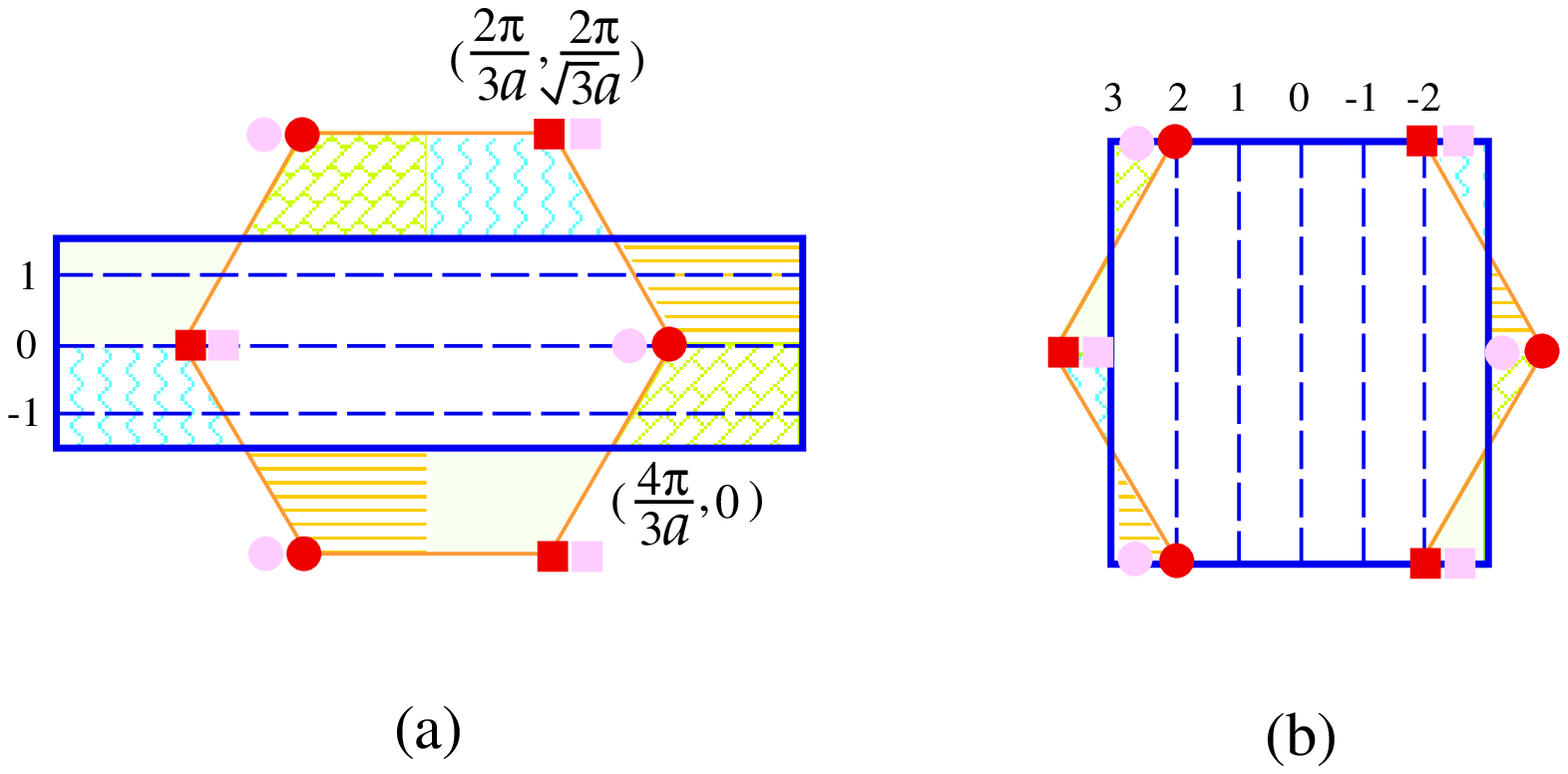}
{\noindent FIG.  2: Dirac points in the Brillouin zone.  Dark circles 
and squares indicate the locations of the Dirac points for 
$t_{\perp}=t$, and grey symbols for $t_{\perp}<t$.  Dashed lines show 
the allowed quantized momenta in (a) the (3,3)-armchair and (b) the 
(6,-6)-zigzag.  The conventional hexagonal zone is mapped into 
rectangle one for (a) the armchair and (b) the zigzag.}
\end{figure}

To proceed with the mapping, the conventional hexagonal BZ needs to be 
rearranged into a rectangle.  With several parts shifted by reciprocal 
lattice vectors, it is done as shown in Fig.  2(a).  It is crucial to 
make such a change and the reason will become clear later.  With the 
choice of this particular BZ, referred to as BZ(arm) later, the 
allowed transverse momentum takes $N_{y}$ integer values
\begin{equation}
k_{y}=\frac{2\pi}{\sqrt{3}a} \frac{p}{N_{y}}, \qquad
p=0, \pm1, \pm2,\ldots, (\pm)[\frac{N_{y}}{2}],
\label{quantized_ky}
\end{equation} 
where $(\pm)$ means, if $N_{y}$ is an even number, only one sign is 
chosen in order to count the total number of modes properly.  For 
notational convenience, we later represent the transverse momentum 
$k_{y}$ by the dimensionless integer $p$.  Similarly, since the 
transverse coordinate $y$ runs through $N_{y}$ values for each fixed 
$x$, it is possible to label it by a dimensionless integer $n$.  
The fermion operator $c_{i}(x,n) \equiv c_{i}(x,y)$ is then labeled by
the integer $n$ defined as
\begin{equation}
n= \left\{ \begin{array}{ll}
\frac{1}{\sqrt{3}a}(y+\delta_{i2} d)&
{\rm if} \hspace{.2cm}x=ma,\\ 
\frac{1}{\sqrt{3}a}(y+\delta_{i2} d)-\frac12&
{\rm if}\hspace{.2cm}x=(m+\frac12)a,
                  \end{array} \right.
\end{equation}
where $m,n$ are integers and $n=1,2,...,N_{y}$. Since only the $k_{y}=0$
mode is important for the armchair at or near half filling, we want to
construct fermion operators carrying definite transverse momentum and then
discard the non-zero momentum components. The required operator is 
constructed by partial Fourier transformation in the $y$-direction
\begin{eqnarray}
c_{i}(x,y) &=& \frac{1}{\sqrt{N_{i}}}
\sum_{k_{x},k_{y}} e^{i{\bf k}\cdot{\bf r}} c_{i}(k_{x}, k_{y})
\nonumber\\
&\equiv&\frac{1}{\sqrt{N_{y}}}
\sum_{p} e^{i\frac{2\pi}{N_{y}}np} d_{i}(x,p),
\label{part_sum}
\end{eqnarray}
where the normalization constant $N_{1}=N_{2}=2N_{y}L/a$ is the total 
number of fermion sites on each sublattice and $L$ is the length of 
the tube.  To obtain the second line in Eq.~\ref{part_sum}, we sum 
over all possible $k_x$ and replace $y$ and $k_{y}$ by integers $n$ 
and $p$ 
respectively.  This transformation can be written down in a more 
compact form
\begin{eqnarray}
c_{i}(x,n)&=& S^{(y)}_{np}d_{i}(x,p),
\label{n_to_p}
\\
d_{i}(x,p)&=& S^{(y)\dag}_{pn}c_{i}(x,n),
\end{eqnarray} 
where summation on doubly appeared indices is implied. The 
transformation matrix $S^{(y)}$ is
\begin{equation}
S^{(y)}_{ab} \equiv \frac{1}{\sqrt{N_{y}}} \exp(i\frac{2\pi}{N_{y}}ab).
\label{yFT}
\end{equation}
This transformation is only possible in the first place because the 
summation over all momenta in the BZ, $\sum_{k_{x},k_{y}}$, can be 
separated into two independent sums, $\sum_{k_{y}} \sum_{k_{x}}$, which 
is a direct result of the rectanglar shape.  Secondly, with the specific 
choice of the BZ(arm), the momentum $k_{y}$ runs through 
$N_{y}$ values (given 
in Eq.~\ref{quantized_ky}) as the transverse coordinate $y$.  As a 
result, the transformation matrix $S^{(y)}$ is unitary which preserves 
the fermionic commutation relations between $d_{i}(x,p)$
\begin{equation}
\{ d_{i}(x,p), d_{j}(x',p') \} = \delta_{ij}\delta_{pp'} 
\delta_{xx'} .
\end{equation} 
Since the Dirac points lie on the $k_{y}=0$ line, all modes with non-zero 
transverse momentum are gapped and can be integrated out in the weak 
coupling. Keeping the zero momentum mode, Eq.~\ref{n_to_p} becomes
\begin{equation}
c_{i}(x,n) \simeq \frac{1}{\sqrt{N_{y}}} d_{i}(x,0)
\equiv \frac{1}{\sqrt{N_{y}}} d_{i}(x),
\end{equation}
where the dependence of the coordinate $y$ completely drops out.
Within this approximation, the Hamiltonian in Eq.~\ref{kinetic}, \ref{on_site}
is mapped to a 2-chain system
\begin{eqnarray}
H_0 =\sum_{x,\alpha}\bigg\{
&-&t d^{\dag}_{1\alpha}(x)d_{2\alpha}(x-b)
\nonumber\\
&-&t d^{\dag}_{1\alpha}(x)d_{2\alpha}(x+b)
\nonumber\\
&-&t_{\perp} d^{\dag}_{1\alpha}(x)d_{2\alpha}(x)+{\rm h.c.} \bigg\},
\label{hop_eff}
\\
H_U= (\frac{U}{N_{y}}) \sum_{x,i} 
&:&d^{\dag}_{i\uparrow}(x)d_{i\uparrow}(x)
d^{\dag}_{i\downarrow}(x)d_{i\downarrow}(x):,
\label{u_eff}
\end{eqnarray}
where $b=a/2$ is the lattice constant.  The hopping term may not look 
familiar at first sight.  By depicting hoppings between two 
sublattices explicitly in Fig.  3(a), it is clear that it describes 
the 
2-chain Hubbard model with lattice constant $b=a/2$.  It is 
interesting that the interaction strength reduces to $U/N_{y}$.  The 
$1/N_{y}$ factor can be thought as coming from delocalized electrons 
around the tube since they would then have a probablility $1/N_{y}$ to 
occupy the same site.  We emphasize that this effective model is only 
correct when the armchair is at or near half filling,
such that the other bands with non-zero transverse momentum are still 
gapped.  Upon larger doping, the interaction effect is more 
complicated and will be studied in the next section.

\begin{figure}[hbt]
\epsfxsize=3.5in\epsfbox{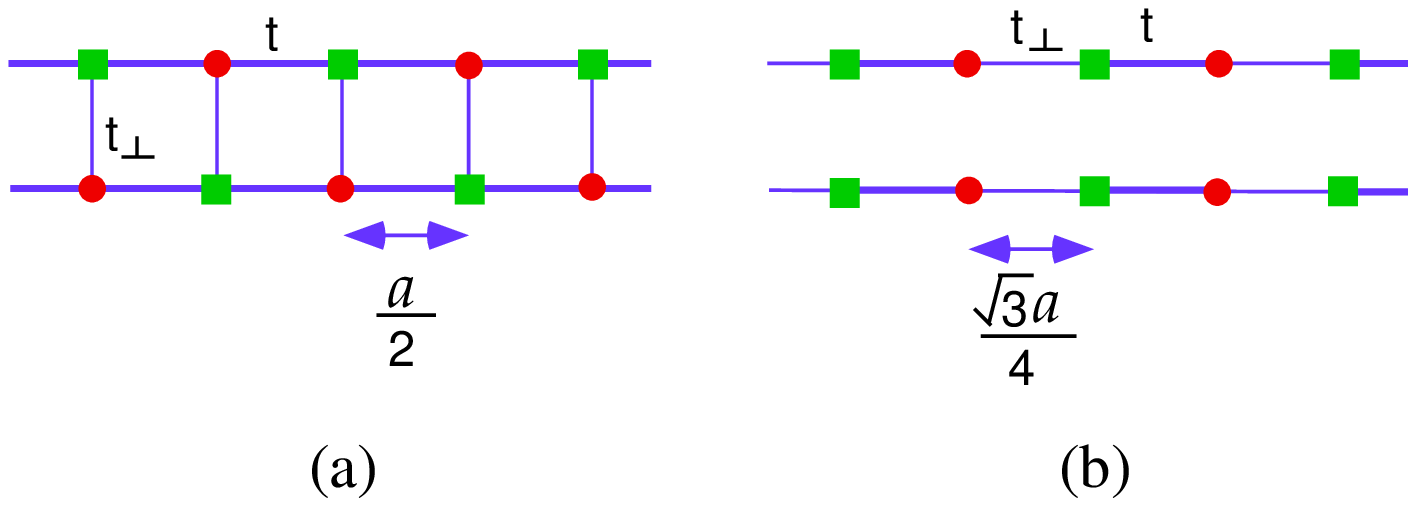}
{\noindent FIG. 3: Effective 2-chain model for (a) the armchair and 
(b) the zigzag. Squres and circles indicate fermion sites on two 
sublattices.}
\end{figure}

The 2-chain Hubbard model has been studied with RG techniques by many 
authors.  Here we adapt the approach and notations in Ref.  
\onlinecite{Lin97u} and briefly review the strategy.  First of all, 
the fermions $d_{i}(x)$ are relabeled by chain indices 
in stead of the sublattice indices 
in Eqs.~\ref{hop_eff}, \ref{u_eff}.  Since the coordinate $x=mb$ is an integer 
multiple of $b$, in the continuous 
limit, the sum is replaced by an integral and the fermion fields acquire 
a rescaling factor
\begin{eqnarray}
b \sum_{x} \to \int dx
\\
\frac{1}{\sqrt{b}}d_{i}(x) = \psi(x,y),
\end{eqnarray}
where $y=1,2$ labels different chains.
We diagonalize the hopping Hamiltonian in the continuous version and 
obtain bonding and anti-bonding bands which are denoted by 
$\psi_{i}(x)$ with $i \equiv k_{y}/\pi=0,1$.  Filling up the bands to the 
chemical potential energy and linearizing the 
dispersion around the Fermi points, 
the fermion fields in bonding and antibonding bands are approximated 
by two pairs of chiral fields as
\begin{equation}
\psi_{i\alpha}(x) \sim \psi^{}_{Ri\alpha}(x) e^{ik_{Fi}x} 
+ \psi^{}_{Li\alpha}(x) e^{-ik_{Fi}x},
\end{equation}
where $k_{Fi}$ is the Fermi momentum of each band. The effective theory
contains two flavors of Dirac fermions described by the Hamiltonian
\begin{equation}
{\cal H}_{0}=\sum_{i\alpha} v_{i}\{ \psi^{\dag}_{Ri\alpha}(x) 
(-i\partial_{x}) \psi^{}_{Ri\alpha} + \psi^{\dag}_{Li\alpha}(x) 
i\partial_{x} \psi^{}_{Li\alpha} \},
\end{equation}
where $v_{i}$ is the Fermi velocity of each band.  There are many possible 
interactions between these Dirac fermions but only some of them are 
relevant to the change of low-energy excitations.  For example, an 
interaction with right (left) moving fields alone only renormalizes 
the Fermi velocities and can be ignored in the leading order of $U$.  
On the other hand, interactions which couple right and left moving 
fields together can possibly open up a gap.  To write down these 
interactions explicitly, the SU(2) scalar and vector currents are introduced
\begin{eqnarray}
J_{ij} &= \psi^{\alpha\dag}_{i} \psi^{\vphantom\dag}_{j\alpha}, 
\hspace{1cm}
\bbox{J}_{ij}&=\frac12 \;\psi^{\alpha\dag}_{i} 
\bbox{\sigma}^{\beta}_{\alpha} \psi^{\vphantom\dag}_{j\beta};
\label{J_def}
\\
I_{ij} &= \psi_{i\alpha} \epsilon^{\alpha\beta} \psi_{j\beta},
\hspace{1cm}
\bbox{I}_{ij}&=\frac12 \;\psi_{i\alpha} \epsilon^{\alpha\beta}
\bbox{\sigma}^{\gamma}_{\beta} \psi_{j\gamma},
\label{I_def}
\end{eqnarray}
where $\bbox{\sigma}$ is Pauli matrices and $\epsilon^{\alpha \beta}$ 
is the Levi-Civita antisymmetric tensor with the convention
$\epsilon^{12}=-\epsilon^{21}=1$. All possible interactions which
are relevant to gap openings are represented succinctly by these currents
\begin{eqnarray} 
{\cal H}_{{\rm int}} &=& c^{\rho}_{ij} J^{R}_{ij} J^{L}_{ij} -
c^{\sigma}_{ij} \bbox{J}^{R}_{ij} \cdot \bbox{J}^{L}_{ij},
\nonumber\\
&+&f^{\rho}_{ij} J^{R}_{ii} J^{L}_{jj} -
f^{\sigma}_{ij} \bbox{J}^{R}_{ii} \cdot \bbox{J}^{L}_{jj},
\label{int_cf}
\\
{\cal H}^{(u)}_{{\rm int}} &=&v^{1\rho}_{ij} I^{R}_{ij} 
I^{L\dag}_{{\hat i}{\hat j}} - v^{1\sigma}_{ij} \bbox{I}^{R}_{ij} \cdot 
\bbox{I}^{L\dag}_{{\hat i}{\hat j}} +{\rm h.c.},
\label{int_u}
\end{eqnarray} 
where $f_{ij}$ and $c_{ij}$ denote the forward and Cooper scattering 
amplitudes respectively, between bands $i$ and $j$.  Umklapp 
interactions $v^{1}_{ij}$\cite{Umklapp:foot}
only appear at half filling.  Summation on 
$i, j$ is implied.  Because of various symmetries, these couplings are 
symmetrical.  Since $\bbox{I}_{ii} \equiv 0$ due to the anti-symmetric
nature of $\bbox{I}_{ij}$, its corresponding coupling $v^{1\sigma}_{ii}$ is 
always set to zero.  Expressing the the on-site interaction in 
Eq.~\ref{u_eff} in terms of these Dirac fermions, we can compare the 
results with Eqs.~\ref{int_cf}, \ref{int_u}.  The initial couplings 
generated by the on-site interaction are
\begin{eqnarray}
c^{\sigma}_{ij}&=&f^{\sigma}_{ij}=4 c^{\rho}_{ij}=4 f^{\rho}_{ij}
= \frac{Ub}{N_{y}},
\label{arm_ini1}
\\
v^{1\rho}_{01} &=&2 v^{1\rho}_{ii}= \frac{Ub}{4N_{y}}, \qquad
v^{1\sigma}_{01}=0,
\label{arm_ini2}
\end{eqnarray}
where $i,j=0,1$ are the band indices.  The lattice constant $b$ 
appears in the continuous limit because these couplings are 
dimensionless.  

With the knowledge of Fermi velocities and initial couplings, we 
employ the RG equations derived in Ref.  \onlinecite{Balents96,Lin97u} 
to determine which couplings are (marginally) relevant and grow into 
the strong coupling regime when short-ranged fluctuations are 
integrated out.  The phase diagram is then determined by these 
relevant couplings through the bosonization technique.  
Numerical results show that, 
at half filling, the relevant couplings gap out 
all spin and charge excitations and lead the system to a C0S0 Mott's 
insulator, where CnSm means there are n gapless charge and m gapless 
spin modes.  The spin and charge gaps change the simple band 
calculation predictions of the transport properties at low temperature.
For instance, the resistivity blows up exponentially when the 
temperature is cooled down below the gap scale. 
Upon doping away from half filling, the umklapp 
interactions are no longer allowed. The Cooper scatterings induce a 
net attrative interaction between electron at low-energy scale and 
lead to electron pairings. According to the numerics,
the system goes into the C1S0 phase with a gap function of d-wave 
symmetry. The spin gap comes from the singlet pairing and 
the only gapless charge mode is the usual U(1) phase in a superconductor. 

As one might notice that the results of the RG analysis only depends 
on two factors: the Fermi velocities calculated from the band 
structure and the initial couplings generated by the on-site 
interaction.  An alternative interpretation of the effective model for 
the armchair is available.  The Fermi velocities can be calculated 
directly from the band structure of the armchair without mapping to 
any effective models.  As to the couplings in Eqs.~\ref{arm_ini1}, 
\ref{arm_ini2} , they are similar to the $N$-chain Hubbard model with 
lattice constant $b$, (e.g.  $c^{\sigma}_{ij}= 2Ub/N$).  By 
comparison, the initial couplings for the $(N_{y},N_{y})$-armchair
in the $N_{f}=2$ region are 
{\em identical} to those of the $2N_{y}$-chain Hubbard model with
two partially filled bands, where $N_{f}$ is the number of Dirac 
fermions in the nanotubes.
It just happens that the band structure for $k_{y}=0$ in the 
armchair has the same dispersion as in a real 2-chain system, which 
enables us to find the mapping described before.  This new 
interpretation has the advantage that the equivalence of the couplings
holds at generic 
fillings, where an effective 2-chain model is impossible.  The 
connection between the $(N_{y},N_{y})$-armchair and the $2N_{y}$-chain 
Hubbard model is elucidated in detail in the next section.

\subsection{Zigzag tubes}
For an $(N_{x},-N_{x})$-zigzag tube, the circumference is $C=N_{x}a$ 
and the momentum $k_x$ is quantized.  The band structure shows 
interesting results even when interactions are negligible.  For now,
we ignore the curvature effect and set $t_{\perp}=t$.  For 
$N_{x} =3n$, the Dirac points lie on the allowed momenta (see Fig.  
2(b)).  The system has gapless excitations and thus is a metal.  For 
$N_{x} \neq 3n$, the allowed momenta do not cut through the Dirac 
points and the (filled) bonding and (empty) antibonding bands are 
separated by a finite gap.  Thus, the system is a semi-conductor.  Due 
to curvature effects, the difference between transverse and 
longitudinal hoppings is of order $1/N_{x}^{2}$ 
\cite{Blase94,Kane97}, and it causes 
a gap $\Delta$ also of order $1/N_{x}^{2}$ even when $N_{x}=3n$.  
However, we 
find later that the effective interaction is of order $1/N_{x}$.  
It is then reasonable to ignore the minor gap caused by 
curvature and treat $(3n,-3n)$-zigzag tubes as metallic.  Since $N_{x} 
\neq 3n$ tubes are already insulating without interactions, the 
inclusion of interaction effect does not change the low-energy physics 
dramatically.  We focus on the more interesting case of 
$N_{x}=3n$ and explore whether the interactions can indeed cause gaps.

For the $(3n,-3n)$-zigzag, a mapping to a 2-chain model, similar to 
the armchair, is found by partial Fourier transform in $x$.  To 
perform the transformation, as for the armchair, a rectanglar BZ is 
again necessary.  However, the BZ(arm) is not suitable because it 
contains $2N_{x}$ quantized $k_{x}$ which is twice the number of modes 
necessary to make the transformation unitary. Another BZ is chosen 
as shown in Fig.  2(b), 
which is referred to as BZ(zz) later.  With this choice, the momentum 
$k_{x}$ has $N_{x}$ allowed values:
\begin{equation}
k_x = \frac{2\pi}{a}\frac{p}{N}, \qquad 
p=0,\pm1,\pm2,...,(\pm)[\frac{N_{x}}{2}].
\label{quantized_kx}
\end{equation}
For each fixed coordinate $y$, the transverse coordinate $x$ runs 
through $N_{x}$
possible values. The fermion field, $c(n,y) \equiv c(x,y)$, is 
labeled by the integer $n$ defined as
\begin{equation}
 n= \left\{ \begin{array}{ll}
                  x/a & {\rm if} \hspace{.2cm}y=m\sqrt{3}a,\\ 
                  x/a -\frac12&{\rm if} \hspace{.2cm}y=(m+\frac12)\sqrt{3}a,
                  \end{array} \right.
\end{equation}
where $m,n$ are integers and $n=1,2,\ldots,N_{x}$.
The fermion field $d_{i}(p,y)$ which carries definite 
momentum $k_{x}$ is constructed by partial Fourier transformation in 
$x$
\begin{eqnarray}
c_{i}(n,y)&=& S^{(x)}_{np}d_{i}(p,y),
\\
d_{i}(p,y)&=& S^{(x)\dag}_{pn}c_{i}(n,y).
\end{eqnarray} 
The transformation matrix $S^{(x)}$ is
\begin{equation}
S^{(x)}_{ab} \equiv \frac{1}{\sqrt{N_{x}}} \exp(i\frac{2\pi}{N_{x}}ab).
\end{equation}
Since $S^{(x)}_{ab}$ is unitary, the fermion field $d_{i}(p,y)$ 
obeys the canonical anti-commutators. For $t_{\perp}=t$, 
the Dirac points lie on two specific momenta,
$k_{x}= \pm 2\pi/3a$, and other bands acquire major gaps. 
Even with the inclusion of curvature effects, the gaps of the
$k_{x}= \pm 2\pi/3a$ bands is only of order $1/N_{x}^{2}$, which is 
smaller than those of the other bands. In the weak coupling, it is 
legitimate to integrate out all other gapped modes and
$c_{i}(x,y)$ only contains components near the Dirac points
\begin{equation}
c_{i}(x,y) \sim \frac{1}{\sqrt{N_{x}}} \sum_{q=\pm} 
d_{qi}(y) e^{iq\frac{2\pi}{3a}x},
\end{equation}
where $d_{\pm i}(y) \equiv d_{i}(\pm N_{x}/3,y)$.  Within this 
approximation, the hopping 
Hamiltonian in Eq.~\ref{kinetic} is simplified to two chains without 
interchain hoppings
\begin{eqnarray}
H_{0} &=& \sum_{y,q=\pm} \bigg\{ -t d^{\dag}_{q1}(y) d^{}_{q2}(y+b'+\delta)
\nonumber\\
&&-t_{\perp} d^{\dag}_{q1}(y) d^{}_{q2}(y-b'+\delta) +{\rm h.c.}\bigg\},
\end{eqnarray}
where $q=\pm$ is the chain indices. 
The lattice constant is $b' = \sqrt{3}a/4$ and the distortion 
distance is $\delta = \sqrt{3}a/12$.  The distortion $\delta$ can be 
gauged away by shifting one of the sublattices.  Performing a gauge 
transformation to shift the second sublattice by $\delta$, we get
\begin{eqnarray}
D^{\dag} &d_{q1}&(y) D = d_{q1}(y),
\\
D^{\dag} &d_{q2}&(y+\delta) D =d_{q2}(y).
\end{eqnarray}
In momentum space, this gauge transformation only adds a phase to
$d_{q2}(k) \to d_{q2}(k) e^{ik\delta}$ and does not change the band
structure at all. As to the interaction, since it is on-site, it is 
also invariant under this gauge transformation. The hopping 
Hamiltonian becomes
\begin{equation}
H_{0}= \sum_{y,q} \{ -t(y) d^{\dag}_{q}(y) d^{}_{q}(y+b')
+{\rm h.c.} \},
\end{equation}
with the hopping $t(y)$ alternating along the $y$-direction as
\begin{equation}
t(y) = \frac12 \big\{ (t+t_{\perp}) +(-1)^{y}(t-t_{\perp}) \big\}.
\end{equation}
For $t_{\perp} \neq t$, the hopping Hamiltonian represents
a dimerized chain as shown in Fig. 3(a)
and has a gap $\Delta \sim (t_{\perp}-t)$ while, 
for $t_{\perp}=t$, the gap vanishes. This is in agreement with our 
previous conclusion for the zigzag. 
As argued before,  the minor 
difference of order $1/N_{x}^{2}$ between $t$, and $t_{\perp}$ caused 
by curvature effects is 
negligible because the interactions are order of $1/N_{x}$. Thus we set 
$t=t_{\perp}$ in the following. 
It takes a little algebra to write 
down the interactions between these two chains and obtain
\begin{eqnarray}
H_{U}&=& (\frac{U}{N_{x}}) \sum_{y,q} \bigg\{ 
n_{q\uparrow}(y) n_{q\downarrow}(y)
+n_{q\uparrow}(y) n_{{\bar q}\downarrow}(y)
\nonumber\\
&+& d^{\dag}_{q\uparrow}(y)d^{}_{{\bar q}\uparrow}(y)
d^{\dag}_{{\bar q}\downarrow}(y)d^{}_{q\downarrow}(y) \bigg\},
\label{int_zz}
\end{eqnarray}
where the notation ${\bar a} =-a$ is adapted through out the paper. 
It is interesting to compare the effective 2-chain models for the 
armchair and the zigzag. For the armchair, the on-site interaction 
remains simple but there exist hoppings between two chains, while, for 
the zigzag, there is no hoppings between chains but the interactions 
are complicated. The effective model obtained here does not look 
like the 2-chain Hubbard model yet. Following the method  for the 
armchair, we take the continuous limit and linearize the spectrum near
the Fermi points. The fermion fields in these two chains are
approximated by two pairs of chiral fermions as
\begin{equation}
\psi_{q}(y) \sim \psi^{}_{Rq}(y)e^{ik_{Fq}y} 
+\psi^{}_{L{\bar q}}(y)e^{-ik_{Fq}y}.
\end{equation}
Once more we have an interacting theory with two flavors of Dirac fermions.  
Because of the absence of interchain hoppings, these two bands are 
degenerate and, thus, always have the same 
Fermi velocities.  The allowed interactions are the same as those of the 
armchairs in Eqs.~\ref{int_cf}, \ref{int_u}.  The initial couplings 
determined from the interactions in Eq.~\ref{int_zz} are
\begin{eqnarray}
c^{\sigma}_{ij}&=&f^{\sigma}_{ij}=4 c^{\rho}_{ij}=4 f^{\rho}_{ij}
= \frac{2Ub'}{N_{x}},
\\
v^{1\rho}_{01} &=&2 v^{1\rho}_{ii}= \frac{Ub'}{2N_{x}}, \qquad
v^{1\sigma}_{01}=0.
\end{eqnarray}
The couplings are exactly the same as for the 2-chain Hubbard 
model with reduced interaction $u_{{\rm eff}}=U/N_{x}$ and 
lattice constant $b'= \sqrt{3}a/4$. 
At half filling, the Fermi velocities of Dirac fermions for 
the armchair and the $(3n,,-3n)$-zigzag are identically the same. As a 
result, the effective theory of the zigzag is again described by the 
2-chain Hubbard model with reduced interaction $U/N_{x}$ and lattice 
constant $b'=\sqrt{3}a/4$. Upon doping, the Fermi velocities of the 
zigzag remain degenerate and start to deviate from the Hubbard model.
However, numerical results show that the phase is still C1S0 as in a 
doped armchair tube. It is interesting to
compare a $(3N,3N)$-armchair and $(6N,-6N)$-zigzag because the 
effective interaction strength $u_{{\rm eff}}=U/3N$ is the same for 
both. However, since the lattice constant 
of the armchair $b$ is slightly larger, the 
initial couplings of the armchair is a bit larger than those of the zigzag. 
By integrating the RG flow numerically, the results imply
that the critical temperature of the 
armchair is higher than the zigzag due to the difference in 
the initial couplings from difference lattice constants.

\section{Nanotubes at generic fillings}
So far, we studied the carbon nanotube at fillings described by
two flavors of interacting Dirac fermions and found it 
is closely related to the 2-chain Hubbard model.  Upon larger doping, 
the chemical potential cuts through more pairs of Fermi points and, 
in general, there are $N_{f}$ flavors (more than two) 
of Dirac fermions participating in the interactions.  
For example, the $(3,3)$-armchair enters $N_{f}=6$ region upon 
larger doping as indicated in Fig.  4, while the $(6,-6)$-zigzag enters 
$N_{f}=4$ region as shown in Fig.  5.  The strategy used to analyze them with 
RG techniques involves two steps: calculate the Fermi velocities 
from the band structure, and figure out all possible interactions with 
the initial couplings generated by the on-site interaction.  
The first step is easy because the band structure of 
nanotubes is already obtained in Eq.~\ref{band}. The second step 
is more challenging. We are going to show that 
the interactions between these $N_{f}$ flavors of Dirac fermions are 
almost the same as in the $N$-chain Hubbard model in the $N_{f}$ 
partially filled band regime.  In the following, we focus on the 
$(3,3)$-armchair in the $N_{f}=6$ region and the $(6,-6)$-zigzag in the 
$N_{f}=4$ region as examples.  The generality of this equivalence is 
transparent from these two typical illustrations.

\begin{figure}[hbt]
\epsfxsize=3.5in\epsfbox{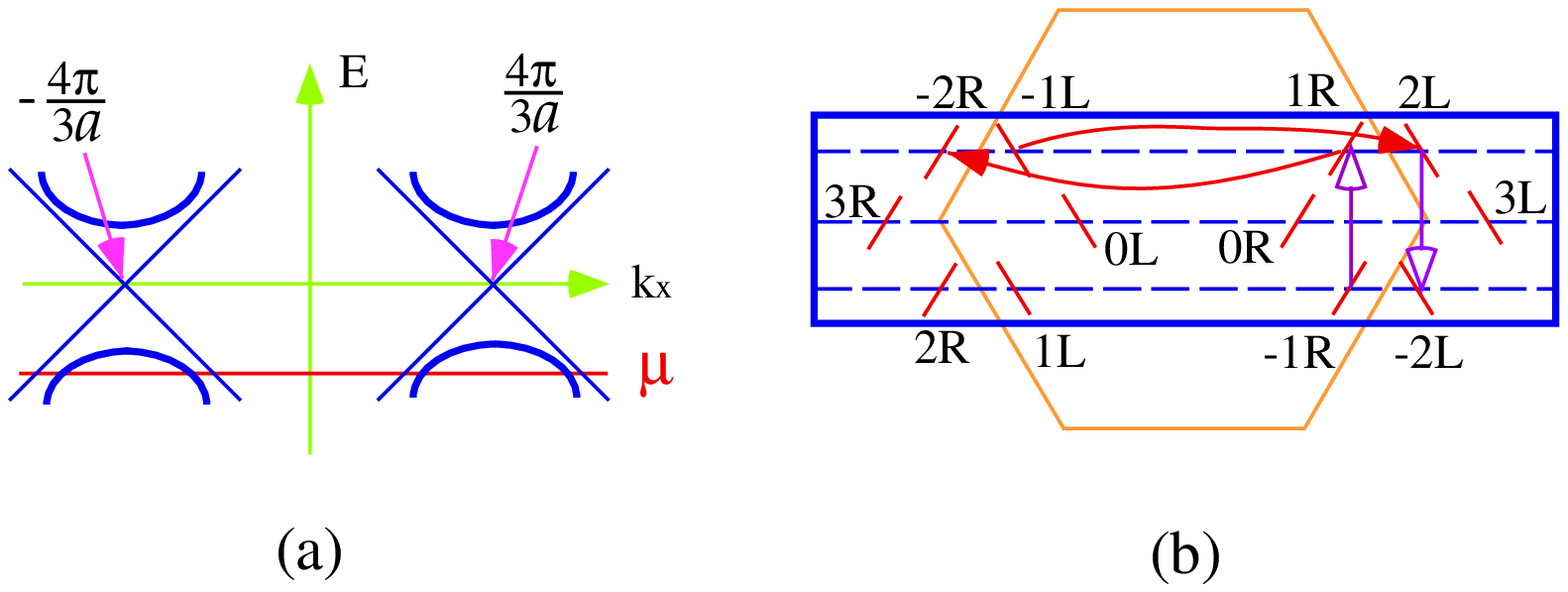}
{\noindent FIG. 4: The (3,3)-armchair in $N_{f}=6$ 
partially filled band region. The band structure near $E=0$ is shown in 
(a) and the corresponding Fermi points are also plotted in the BZ in (b). 
The thick line represents the two degenerate bands with transverse 
momentum $p=\pm 1$.
For each quantized momentum $p$, there are two flavors of Dirac 
fermions participating in the interactions. The solid (empty) arrow lines 
indicate the umklapp coupling $u^{1}_{1{\bar 2}}$ ($u^{2}_{1{\bar 
2}}$) in addition to the Cooper and forward couplings.}
\end{figure}

We study the armchair first. 
For convenience, the sublattice index $i=1,2$ in $c_{i}(x,y)$ is viewed 
as the $z$-coordinate. The corresponding momentum $p' \equiv 
k_{z}/\pi=0, 1$ labels the bonding and anti-bonding bands respectively.
Perform a partial Fourier transformation in BZ(arm)
for both $y, z$ to diagonalize the hoppings around the tube, we get
\begin{equation}
c_{i}(x,n) = S^{(y)}_{np} S^{(z)}_{ip'} c_{{\bf p}}(x),
\end{equation}
where ${\bf p} = (p, p')$ is the momentum in the $y, z$ directions.
The transfer matrix $S^{(z)}$ is defined analogous to $S^{(y)}$ 
defined in Eq.~\ref{yFT}
\begin{equation}
S^{(z)}_{ab} = \frac{1}{\sqrt{N_{z}}} \exp (i\frac{2\pi}{N_{z}} ab),
\end{equation}
where the normalization constant $N_{z}=2$ accounts 
for two possible values of $z (k_{z})$.  
In the continuous limit, the fermion field is rescaled as
\begin{equation}
c_{{\bf p}}(x) = \sqrt{b} \psi_{{\bf p}}(x).
\end{equation}
We linearize the dispersions near each Fermi point and describe
the system by $N_{f}=6$ flavors of Dirac fermions. Since the system is 
doped, only the bonding band ($p'=0$) participates in the interactions.
Since the anti-bonding band ($p'=1$) never appear in the following
analysis, the $p'$ index is dropped for notational simplicity.
To make the equivalence to the 
Hubbard model transparent, the fermions are labeled in a particular 
way:
\begin{eqnarray}
\psi_{p}(x) &\sim &\psi^{}_{Rp}(x) e^{ik_{Fp}x}
+\psi^{}_{L{\bar p}}(x) e^{-ik_{Fp}x}
\nonumber\\
&+&\psi^{}_{Rp^{*}}(x) e^{ik_{Fp^{*}}x}
+\psi^{}_{L{\bar pp^{*}}}(x) e^{-ik_{Fpp^{*}}x},
\end{eqnarray}
where $p^{*}=p-{\rm sgn}(p) N_{y}$. 
The labeling of these fermions is given explicitly in the Fig. 4.  
It is straightforward to check that the Cooper and 
forward scatterings are allowed between any two of the fermions. 
However, since there are two flavors of Dirac fermions for each
$k_{y}$, additional interactions are also allowed. Two examples are 
shown in Fig 4(b). These are 
the umklapp interactions in the $k_{y}$ direction
\begin{eqnarray} 
{\cal H}^{(uy)}_{{\rm int}} &=& u^{1\rho}_{ij}J^{R}_{ij} J^{L}_{{\bar i}{\bar j}} 
-u^{1\sigma}_{ij}\bbox{J}^{R}_{ij} \cdot \bbox{J}^{L}_{{\bar i}{\bar j}},
\nonumber\\
&+&u^{2\rho}_{ij} J^{R}_{i{\bar i}} J^{L}_{j{\bar j}} -
u^{2\sigma}_{ij} \bbox{J}^{R}_{i{\bar i}} \cdot \bbox{J}^{L}_{j{\bar j}}.
\label{umk_ky}
\end{eqnarray}
The name comes from the fact that if the BZ(arm) is shifted back to the 
the conventional one, the transfer transverse momentum associated
with each vertex equals the reciprocal lattice vector 
$4\pi/\sqrt{3}a $. The 
allowed interactions are therefore described by Eqs.~\ref{int_cf}, 
\ref{umk_ky} as in a doped (even-chain) Hubbard model with 
periodic boundary 
conditions. We need to compute the initial couplings generated by 
the on-site interaction to complete the equivalence. The on-site 
interaction is expressed in terms of the Dirac feermions as
\begin{eqnarray}
H_{U}&=&  U \sum_{x,y,z} n_{\uparrow}(x,y,z) n_{\downarrow}(x,y,z)
\nonumber\\
&=& Ub \int dx \sum_{P_{i_{a}}, i_{a}} \bigg\{
f^{(y)}_{i_{1}i_{2}i_{3}i_{4}} f^{(z)}_{j_{1}j_{2}j_{3}j_{4}}
\nonumber\\
&&\psi^{\dag}_{P_{i_{1}}i_{1}\uparrow}
\psi^{}_{P_{i_{2}}i_{2}\uparrow}
\psi^{\dag}_{P_{i_{3}}i_{3}\downarrow}
\psi^{}_{P_{i_{4}}i_{4}\downarrow}
e^{iQ_{x}x} \bigg\},
\end{eqnarray}
where $i_{a}$ is the momentum index of $k_{y}$ and run through
$N_{f}=6$ values. The $j_{a}$ labels the $k_{z}$ momentum and is always
zero because only the boning band is interactiing.
The chirality of the fermions are denoted by $P_{i_{a}} = R, L$.
The phase associated with each vertex is 
the transfer momentum in $x$-direction given by,
\begin{equation}
Q_{x} \equiv -P_{i_{1}} k_{Fi_{1}} +P_{i_{2}} k_{Fi_{2}}
-P_{i_{3}} k_{Fi_{3}}+P_{i_{4}} k_{Fi_{4}}.
\end{equation}
Conservation of crystal momentum requires $Q_{x}$ equal to one of 
the reciprocal lattice vectors. If the momentum is not conserved, the 
phase associated with the vertex oscillates and the vertex is 
irrelevant in the leading order analysis.
The summations of the transformation matrices
in the $y, z$ directions are
\begin{eqnarray}
f^{(d)}_{i_{1}i_{2}i_{3}i_{4}} \equiv \sum_{l}
S^{(d)*}_{li_{1}}S^{(d)}_{li_{2}}S^{(d)*}_{li_{3}}S^{(d)}_{li_{4}}
=\frac{1}{N_{d}} \delta_{Q_{d}, G_{d}},
\end{eqnarray}
where $d = y, z$ and $Q_{d}=(-i_{1}+i_{2}-i_{3}+i_{4})$ is the 
transfer momentum and $G_{d}=2nN_{d}$ is the reciprocal lattice vector.
The $\delta$-function enforces the crystal 
momentum conservation requirement.
Putting everything together, the interacting 
Hamiltonian, in the continuous limit, is
\begin{equation}
H_{U}= (\frac{Ub}{2N_{y}}) \int dx
\psi^{\dag}_{P_{i_{1}}i_{1}\uparrow}
\psi^{}_{P_{i_{2}}i_{2}\uparrow}
\psi^{\dag}_{P_{i_{3}}i_{3}\downarrow}
\psi^{}_{P_{i_{4}}i_{4}\downarrow},
\end{equation}
where summation over all possible vertices $P_{i_a}, i_{a}$, 
constrainted by momentum conservation, is implied. 
These allowed vertices which can be cast into current-current
interactions in Eqs.~\ref{int_cf}, \ref{umk_ky}, have been discussed
before. The key point here is that 
the prefactor $Ub/2N_{y}$ obtained above is 
indeed {\em exactly} the same as that of the $2N_{y}$-chain 
Hubbard model with lattice constant $b=a/2$ in the $N_{f}$ filled 
band region. Therefore the equivalence of interactions between the 
$(N_{y},N_{y})$-armchair and the $2N_{y}$-chain Hubbard model becomes
clear.

Knowing the above equivalence, the (3,3)-armchair in the 
$N_{f}=6$ region can be studied 
by integrating the RG equations numerically. Two of the Dirac 
fermions at $k_{y}=0$ become completely decoupled and contribute two 
gapless charge and two gapless spin modes (C2S2). One recalls that 
these two Dirac fermions form d-wave pairs and result in a C1S0 
superconducting phase in the $N_{f}=2$ region.   
Relevant interactions between the 
remaining four flavors of fermions are too complicated to be 
bosonized. Therefore, the phase of the armchair in 
the $N_{f}=6$ region remains an open question for future study.

\begin{figure}[hbt]
\epsfxsize=3.5in\epsfbox{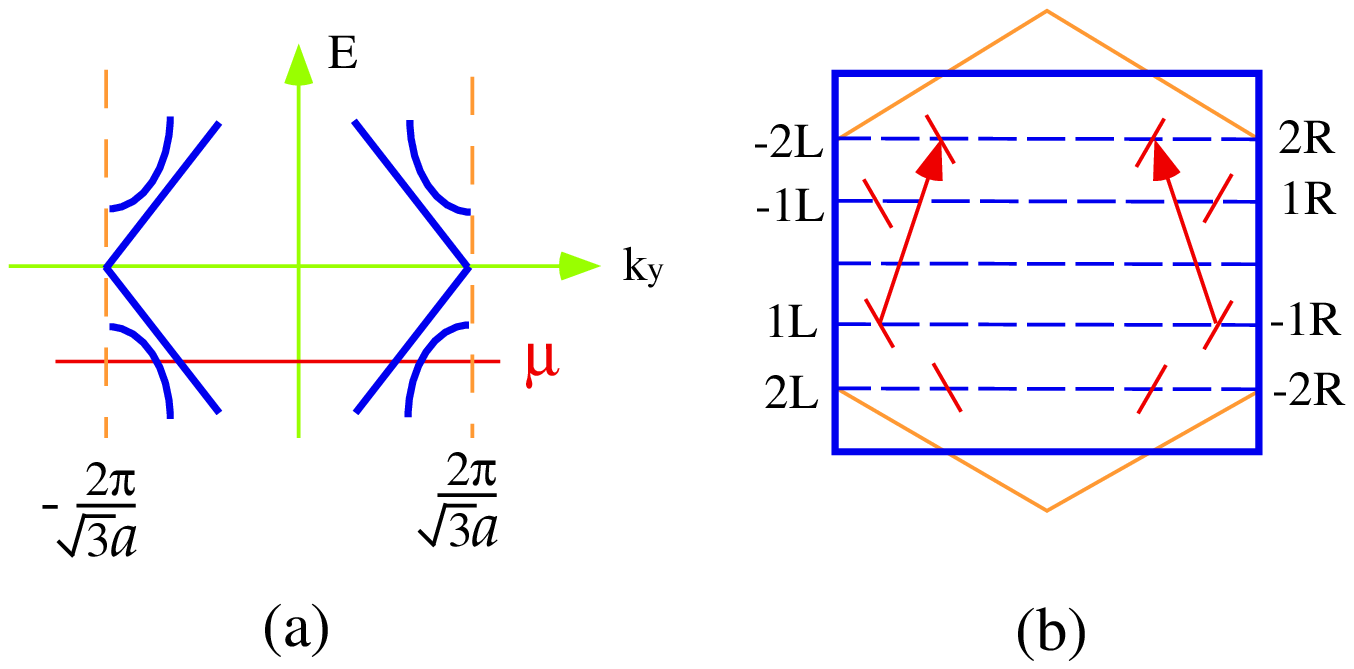}
{\noindent FIG. 5: The (6,-6)-zigzag in $N_{f}=4$ partially filled 
band region. The band structure near $E=0$ is shown in 
(a) and the corresponding Fermi points are also plotted in the BZ in (b). 
The interaction between Band $\pm 1, \pm 2$, as indicated by solid 
lines, is not allowed because the momentum transfer is {\em not} a
reciprocal lattice vector.}
\end{figure}

Now we turn to the zigzag.
Perform a partial Fourier transformation in both $x, z$ to diagonalize
the hoppings around the tube:
\begin{equation}
c(y,m',n') = S^{(x)}_{m'm} S^{(z)}_{n'n} c_{{\bf p}}(y),
\end{equation}
Since the coordinate $y=2nb'$ is an integer multiple of $2b'$, 
in the continuous limit, the fermion field is rescaled as
\begin{equation}
c_{{\bf p}}(y) = \sqrt{2b'} \psi_{{\bf p}}(y).
\end{equation}
Linearizing the spectrum near the Fermi points 
results in following four flavors of Dirac fermions
\begin{eqnarray}
\psi_{p}(x) \sim &\psi^{}_{Rp}&(x) e^{ik_{Fp}x}
+\psi^{}_{L{\bar p}}(x) e^{-ik_{Fp}x}.
\end{eqnarray}
The Cooper and forward scatterings  are again allowed because the 
transfer momenta of these vertices are conserved.
The $k_{y}$ umklapp interactions are not allowed because the Fermi 
momenta do not sum up to the reciprocal lattice vector as shown in Fig. 
5. The initial couplings of these interactions are
\begin{equation}
H_{U}= (\frac{Ub'}{N_{x}}) \int dy
\psi^{\dag}_{P_{i_{1}}i_{1}\uparrow}
\psi^{}_{P_{i_{2}}i_{2}\uparrow}
\psi^{\dag}_{P_{i_{3}}i_{3}\downarrow}
\psi^{}_{P_{i_{4}}i_{4}\downarrow}.
\end{equation}
This interacting Hamiltonian is the same as the $N_{x}$-chain Hubbard 
model with lattice constant $b'$ in the $N_{f}$ partially filled band 
region, except that the $k_{y}$ umklapp interactions are missing.

With this equivalence, the (6,-6)-zigzag may be studied as follows.  
Upon doping into the $N_{f}=4$ region, the superconductivity vanishes.  
The bands near the Dirac points are decoupled and contribute 2 gapless 
charge and 2 spin excitations (C2S2).  The remaining two bands form 
the usual ``d-wave'' pairing and contribute C1S0.  The final phase 
then is C3S2 and the system no longer has a spin gap. This result 
shows that the superconductivity only exists inside a narrow window
of doping for these nanotubes.

\section{Discussions and Conclusions}
Experimental studies of multi-wall nanotubes show that the electrical 
properties vary from tube to tube, and are dominated by disorder 
and weak localization in stead of the tubule structure\cite{Dai96,Ebbesen96b}.
The single-wall nanotubes (SWNT's) on the other hand have simpler 
structures and are expected to be more helpful to clear up various 
interaction effects.  Recently, they have been synthesized with high 
yields and structure uniformity\cite{Thess96}.  The (10,10)-armchairs
dominate these bulk samples\cite{Cowley97}.
Tubes during the production process 
form a 2D triangular lattice (crystalline ropes) and eventually become 
mats (three dimensional sample with entangled ropes).  The 
(10,10)-armchair is predicted to be metallic (within band calculation) 
in agreement with some of the experimental evidence.  
Some of the transport measurements show 
that the resistivity of the ropes and mats increases linearly with 
temperature in the high temperature regime, while at low temperature,
one observes a crossover to negative 
$d\rho/dT$\cite{Fischer97}. 
However, the experimental results are not yet conclusive.
From purely electronic 
(Coulomb) interactions, it was shown that the umklapp interactions at 
half filling cause a linear increase of resistivity with temperature
in the region above the Mott insulating gap, and an exponential increase
upon cooling below the gap\cite{Balents97}.
Another approach explains the linear
resistivity at high temperature by the study of thermal shape 
fluctuations of these tubes\cite{Kane97,Kane97u}.
These two approaches produce the same predictions at high temperature.
However, at low temperature, the resistivity caused by phonons
still decreases linearly with temperature in contrast to the 
exponential increase due to umklapp interactions. A precise 
resistivity measurement at low temperature is necessary to distinguish 
these two contributions.
However, the observed 
dependence of this crossover on the samples' morphology and quality 
suggests that disorder or (and) couplings between single tubes may dominate 
the low temperature behavior.  Because the present experimental 
results are not yet clear in this regime, it will be important to 
carry out more experiments to understand the origin of this turnover 
in the resistivity measurements.

Upon doping, both of the $(N,N)$-armchair and $(3N,-3N)$-zigzag with 
on-site interactions become superconductors (SC's) as discussed in the 
text. Recent studies show that this SC phase is stable with inclusion of
moderate nearest neighbor interactions\cite{Lin97u2}.
It implies that the SC phase 
probably always exists for generic short-range interactions. So far, there
is no theoretical work which includes the long-range interaction effect.
It would be interesting to see how the unscreened Coulomb repulsion 
affects this superconducting behavior in the future study.

In conclusion, we represent nanotubes at generic fillings as $N_{f}$ 
flavors of interacting Dirac fermions whose velocities are determined 
by the band structures.  With specific on-site type interactions for 
the $(N_{y},N_{y})$-armchair and the $(N_{x},-N_{x})$-zigzag, we found 
the vertices equivalent to those of the $2N_{y}$- and $N_{x}$-chain 
Hubbard model respectively, which can be analyzed by the RG method in 
the weak coupling.  With this equivalence, mappings from nanotubes to 
the effective 2-chain model with reduced couplings are achieved at or 
near half filling.  The (3,3)-armchair and the (6,-6)-zigzag are 
studied and compared as an example because both of them 
resemble 6-chain systems. At half filling, they are both Mott's 
insulators and become superconductors upon slight doping. A larger 
doping to $N_{f}>2$ regime destroys the superconductivity.

\acknowledgements We are grateful to Matthew P. A. Fisher for 
motivating the work and many fruitful discussions. We also thank Leon 
Balents for helpful conversations.
This work has been supported by the National Science
Foundation under grant Nos. PHY-9407194, DMR-9400142, DMR-9528578.

\end{multicols}
\end{document}